\documentclass[11pt]{elsarticle}

\usepackage{amsmath}
\usepackage{here}
\pdfoutput=1

\makeatletter
\def\ps@pprintTitle{%
  \let\@oddhead\@empty
  \let\@evenhead\@empty
  \let\@oddfoot\@empty
  \let\@evenfoot\@oddfoot
}
\makeatother

\usepackage{url}
\usepackage{breakurl}
\usepackage[breaklinks,
            colorlinks = true,
            linkcolor = blue,
            urlcolor  = blue,
            citecolor = blue,
            anchorcolor = blue]{hyperref}

\usepackage{graphicx}
\usepackage[margin=1.25in]{geometry}
\usepackage[usenames,dvipsnames]{color}

\newcommand\acp{\begin{center}
\rule[-0.2in]{\hsize}{0.01in}\\\rule{\hsize}{0.01in}\\
\vskip 0.1in 
The African Conference on Fundamental and Applied Physics
    \vskip 0.05in
    {\it Second Edition, ACP2021, March 7--11, 2022 --- Virtual Event}\\
\rule{\hsize}{0.01in}\\\rule[+0.2in]{\hsize}{0.01in} \\
\end{center}}

\usepackage[firstpage=true]{background}
\backgroundsetup{contents={\parbox{6.5in}{\acp}}, scale=1,placement=top,opacity=1,color=black,position={3.25in,1.2in}}

\usepackage{fancyhdr}
\fancypagestyle{plain}{%
  \fancyhf{}%
  \fancyhead[C]{}
  \fancyfoot[C]{\thepage}
}

\fancypagestyle{empty}{%
  \fancyhf{}%
  \fancyhead[C]{{\it ACP2021, March 7--11, 2022 --- Virtual Edition}}
  \fancyfoot[C]{\thepage}
}
\begin{document}

\begin{frontmatter}

\title{Demystifying the fractional order phase transitions  in AdS black holes}

\author[add1]{Mohamed Chabab}
\ead{mchabab@uca.ac.ma}
\address[add1]{High Energy and Astrophysics Laboratory, FSSM, Cadi Ayyad University, P.O.B. 2390  Marrakech, Morocco}

\author[add1,add2]{Samir Iraoui\corref{cor1}}
\cortext[cor1]{Corresponding Author}
\ead{samir.iraoui@ced.uca.ma}
\address[add2]{	Regional Center for Education and Training, Marrakech, Morocco}

\begin{abstract}
\noindent 
In this work, we performed a detailed study of the fractional order phase transition (FPT) for several AdS black hole prototypes~\cite{Chabab:2020imt}.
Our objective is to see whether  the FPT $4/3$ order  at critical points reported in \cite{Ma:2019hvr} is universal. Our analysis shows two remarkable features:  Firstly,  the FPT  is located at $4/3$ order only when the black hole possesses a spherical symmetry. Secondly, this fractional order is not universal and can be altered  by the geometric symmetry.  
\end{abstract}

\begin{keyword}
	AdS  black holes\sep thermodynamics\sep Gauss-Bonnet\sep Kerr black holes\sep  fractional order phase transition.
	
\end{keyword}

\end{frontmatter}

\section{Introduction}
\label{sec:intro}
\noindent
Many remarkable features in the black hole systems were uncovered in the oustanding work of Hawking and Page~\cite{Hawking:1982dh}. Particularly, it has been shown that Schwarzschild AdS black holes are  in stable equilibrium with the thermal AdS space. Besides,  the phase transition between thermal radiation states and  stable large black holes (\textbf{BH}) can take place at a certain critical temperature. This phenomenon is known as Hawking-Page phase transition.

Recently, the study of  BH thermodynamics  has been generalized to the extended phase space, where the cosmological constant is identified as a thermodynamic pressure and treated as a variable, while its conjugate quantity is considered as the thermodynamic volume~\cite{Kubiznak:2012wp}.
Furthermore,  the inclusion of the $P-V$ term in the first law of thermodynamics has led to identification of the  black hole mass with the enthalpy of the event horizon~\cite{Kastor:2009wy}. Then, the analogy with the Van de Waals $P-V$  criticality has been established~\cite{Dolan:2011xt} and the first and second order phase transitions readily found~\cite{Kubiznak:2012wp}.  Indeed, the electric charge plays a primordial role which  gives rise to a small/large BH phase transition analogous to that usually found in ordinary thermodynamic systems~\cite{Dolan:2011xt}. Since then, many similar thermodynamical studies for other solutions of black holes have been performed~\cite{Gunasekaran:2012dq,2014, Chabab:2020imt}. 

Furthermore, to probe and classify thermodynamic phase transitions, Ehrenfest used the discontinuities in derivatives of the free energy. Ehrenfest classification is generally based on the following principle:  phase transition is of order $n$ if the free  energy function and its derivatives up to order $n-1$ are continuous, while, at least, one of its derivatives of order $n$  is discontinuous at critical point. Hawking-Page phase transitions is an example of first order transition. 

A more general classification scheme relying on fractional derivatives, i.e. a derivative with non-integer order in the thermodynamic state space, has been proposed~\cite{1991,1992,2000}.  Fractional phase transitions (FPT) go back to the work of Nagle on dipalmitoyl lecithin (DPL) system~\cite{1973},  where it has been observed  that the order/disorder phase transition is of  $3/2$ order.  The first  generalization to AdS black holes was recently conducted in ~\cite{Ma:2019hvr} where it has been shown that RN-AdS black holes can undergo FPT of order $4/3$.

In the work,  we will explore the fractional phase transitions of several AdS black hole  spherical  solutions, with the objective  to see if fractional $4/3$ order still holds for any AdS black hole prototypes.
 Section~\ref{Kerr} will be devoted to Kerr-AdS black hole and the analysis of its FPT order. This allows us to show the effect of geometric symmetry on FPT. Our conclusion will be drawn in the last section.

\section{Thermodynamics  of AdS spherical black holes and fractional order phase transition}
\label{Thermo}
Herer we consider  three spherical symmetric solutions of static asymptotically  AdS black holes, namely: a charged-AdS BH surrounded by quintessence, $D$ dimensional RN-AdS BH and a $5$D Gauss-Bonnet AdS black holes, given respectively by:
\begin{align}\label{Kilesev}
f(r)_{Quint}&=1-\frac{2 M}{r}+\frac{Q^{2}}{r^{2}}-\frac{ \Lambda}{3}r^{2}-\frac{\alpha}{r^{3\omega_{q}+1}},\\
f_D(r)&=1-\frac{m}{r^{D-3}} +\frac{q^2}{r^{2(D-3)}}    -\frac{2 \Lambda  r^2}{(D-2) (D-1)},\\
f_{GB}(r)&=1+\frac{r^{2}}{2 \tilde{\alpha}}\left(1-\sqrt{1-\frac{16 \pi P\tilde{\alpha}}{3}+\frac{32 M \tilde{\alpha}}{3 \pi r^{4}}}\right),
\end{align} 
where $\alpha$ represents a positive normalisation parameter,  $M$ and $Q$ are the mass  and  electric charge of the black hole, $\tilde{\alpha}$ is a  Gauss-Bonnet coupling constant and $\omega_q$ is the quintessential state parameter. 
While the parameters $m$ and $q$     are related to the  mass and charge of black holes by~\cite{Chamblin:1999tk}
\begin{align}
M=\frac{D-2}{16 \pi} \Omega_{D-2} m,\qquad Q=\frac{\sqrt{2(D-2)(D-3)}}{8\pi}\,\Omega_{D-2}\,q,
\end{align}
where $\Omega_D$ is the volume of the unit $D$-sphere.

As usual, we treat the cosmological constant as a dynamical pressure of the black hole~\cite{Kastor:2009wy} by the relation $P=-\frac{\Lambda}{8\pi}$.
The Hawking temperature  related to the surface gravity via the formula $2\pi T=\kappa$
~\cite{Chabab:2020ejk}.

For subsequent analysis, it is more appropriate to introduce  the following dimensionless variables,
\begin{equation}\label{dimpvt}
p=\frac{P}{P_c}-1, \qquad t=\frac{T}{T_c}-1,  \qquad \nu=\frac{v}{v_c}-1,
\end{equation}
where $P_c, v_c, T_c$ are the coordinates of the critical point.
Thus with the new set of variables $({t}, {p}, {\nu})$, the critical point is located at $({t}={p}={\nu}=0)$.
We can now obtain the dimensionless Gibbs free energy for each black hole solution:
\begin{itemize}
	\item 
Charged-AdS BH surrounded by quintessence 
\begin{equation}g(t,p)=\frac{8-\nu ^4-4 \nu ^3+8 \nu -(\nu +1)^4 p}{4 \sqrt{6} (\nu +1)};
\end{equation}
\end{itemize}
\begin{itemize}
\item
Charged AdS black holes in higher dimensions
\end{itemize}
{\footnotesize\begin{align}
g(t,p)=\frac{(5-2 D)^2 (\nu +1)^{3-D}}{2 [(D-2) (2 D-5)]^{3/2}}+\frac{1}{2} \sqrt{\frac{2 D-5}{D-2}} (\nu +1)^{D-3}-\sqrt{\frac{2 D-5}{(D-2)^3}}\, \frac{(D-3)^2  }{2
	(D-1)}(p+1) (\nu +1)^{D-1};
\end{align}}
\begin{itemize}
\item
$5D$ Gauss-Bonnet-AdS black hole 
\begin{equation}
g(t,p)=\frac{\nu ^3 (\nu +2)^3+(\nu  (\nu +2)+4) (\nu +1)^4 p}{8 (3 \nu  (\nu +2)+4)}.
\end{equation}
\end{itemize}

Next we probe  the behavior of  $g(t,p)$ and its fractional derivatives near critical point. 
We want to express $g(t, p)$ in the Taylor series of $t$, and written in a simple form as,
\begin{equation}
g(t, p)=A_{0}( p)+A_{1}( p) t+A_{2}( p) t^2+\dots
\end{equation}

Here we use  Caputo's definition of fractional derivatives~\cite{caputo}:
\begin{equation}
D_{ t}^{\beta}g( t)=\frac{1}{\Gamma\left(n-\beta\right)}\int_{0}^{t}\left(t-\tau\right)^{n-\beta-1}\, \frac{\partial^{n}g(\tau)}{\partial \tau^{n}}\mathrm{d}\tau,\qquad  n-1<\beta<n,
\end{equation}
where $\beta$ is the order of derivative and  $n$  an integer. 
For RN-AdS BH surrounded by quintessence, we can calculate the values of $D_{ t}^{\beta}g( t, p)$ for $1<\beta<2$ in the limit $( t \rightarrow 0,~ p\rightarrow 0)$. 
Thus, we  obtain
\begin{equation}
\lim_{ t\rightarrow 0^{\pm}}D_{ t}^{\beta}g( t, p)=\left\{
\begin{array}{lr}
0 ~~&\text{for}~~\beta<4/3,\\
\mp\frac{\left(2-3 \sqrt{6} \alpha  Q\right)^{4/3}}{\sqrt{6} \Gamma \left(\frac{2}{3}\right)} &\text{for}~~\beta=4/3,\\
\mp\infty ~~ &\text{for}~~\beta>4/3.
\end{array}
\right.\end{equation}
Clearly, for  $\beta=4/3$, we see a jump discontinuity,
$
\lim_{ t\rightarrow 0^{-}}D_{ t}^{\beta}g \neq \lim_{ t\rightarrow 0^{+}}D_{ t}^{\beta}g.
$
When $\beta > 4/3$, the $\beta$-order fractional derivatives of the Gibbs free energy diverge. Hence, as for $4D$ RN-AdS black holes~\cite{Ma:2019hvr}, the FPT order of the RN-AdS Black hole surrounded by quintessence is $\beta=4/3$. 
Same result is obtained for the other two prototypes. Indeed, we find  for 5D Gauss-Bonnet-AdS black hole that the phase transition near the critical point is fractional  and happens at order 4/3. Likewise,  whatever the dimension $D$ of  the RN-AdS black hole, the order of FPT  arises decidedly at $\beta=\frac{4}{3}$.
We summarized the  calculation results for all dimensions $D=4 -10$ in Table~\ref{tableRNAdS}.

\begin{table}[!h]
	\begin{center}
		\begin{tabular}{ccccc}
			\hline 
			Dimension $D$& $D=4$  &  $D=5$ & $D=6$  & $D=7$  \\  
			order FPT &  $\beta=4/3$ & $\beta=4/3$  &  $\beta=4/3$ & $\beta=4/3$  \\  
			$\lim_{ t\rightarrow 0^{\pm}}D_{ t}^{\beta}g( t, p)$ & $\mp \dfrac{2\ 2^{5/6}}{3 \sqrt{3}\, \Gamma \left(\frac{5}{3}\right)}$   & $\mp \dfrac{8 \sqrt[3]{2}}{\sqrt[6]{3}\, 5^{5/6}\, \Gamma \left(\frac{5}{3}\right)}$  &  $\mp  \dfrac{12 \sqrt[3]{6}}{7^{5/6} \, \Gamma \left(\frac{5}{3}\right)}  $ & $\mp  \dfrac{32 \sqrt[3]{\frac{2}{3}} \sqrt{5}}{9\, \Gamma \left(\frac{5}{3}\right)}  $    \\
			\hline 
		\end{tabular} 
	
	\vspace*{.5em}
	\begin{tabular}{ccccc}
		\hline 
		Dimension $D$&  $D=7$  & $D=8$  &  $D=9$&  $D=10$\\  
		order FPT &   $\beta=4/3$  & $\beta=4/3$ & $\beta=4/3$ & $\beta=4/3$\\  
		$\lim_{ t\rightarrow 0^{\pm}}D_{ t}^{\beta}g( t, p)$ &  $\mp  \dfrac{32 \sqrt[3]{\frac{2}{3}} \sqrt{5}}{9\, \Gamma \left(\frac{5}{3}\right)}  $   & $\mp  \dfrac{50 \left(\frac{2}{11}\right)^{5/6}}{\sqrt[6]{3}\, \Gamma \left(\frac{5}{3}\right)}  $ &  $\mp  \dfrac{24 \sqrt[3]{6} \sqrt{7}}{13^{5/6} \, \Gamma \left(\frac{5}{3}\right)}  $ & $\mp  \frac{196 \left(\frac{2}{5}\right)^{5/6}}{3 \sqrt{3} \, \Gamma \left(\frac{5}{3}\right)}  $ \\
		\hline 
	\end{tabular} 
		\caption{Limits and fractional order phase transition at the critical point for $D$ dimensional RN-AdS BH.}
		\label{tableRNAdS}
	\end{center}
\end{table}

This  suggests that FPT is inevitably located at $4/3$ order as long as the AdS black hole solution possesses a spherical symmetry and neither the spacetime dimension nor the higher derivative corrections introduced by means of the Gauss-Bonnet term can alter the fractional order  phase transition at the critical point.  Even the external quintessence field surrounding the charged AdS black hole does not affect this order.

\section{Fractional order phase transition of Kerr-AdS black holes}
\label{Kerr}

The  AdS rotating black hole solution is given by the Kerr-AdS metric~\cite{Kerr:1963ud}.
Using this metric, we can calculate the expression of rescaled Gibbs free energy  in terms of the dimensionless parameters given in Eq.~\eqref{dimpvt}:
\begin{align}\label{Gibbsrescaled}
g(t,p)=G/\sqrt{J}=\frac{\sqrt[4]{5} \left(-12 \sqrt[4]{10} \sqrt{J} (\nu +1)^7 (p+1)+9 \sqrt{3} (\nu
	+1)^4+\sqrt{3}\right)}{18\ 2^{3/4} (\nu +1)^3}.
\end{align}

One can readily  derive the phase transition order near the critical point. Indeed,  the t-expansion of the rescaled Gibbs free energy becomes, 
{\footnotesize\begin{multline}\label{gKerr}
g(t,p)\approx C_1
+\left[C_2+\frac{C_3}{p^{2/3}}+\frac{C_4}{p^{1/3}}+\dots\right]t  
+\left[C_5+\frac{C_6}{p^{5/3}}+\frac{C_7}{p^{4/3}}+\frac{C_8}{p^{2/3}}++\frac{C_9}{p^{1/3}}+\dots\right]t^2 + \mathcal{O}[ t^3].
\end{multline}}
Using the critical point calculated in~\cite{Gunasekaran:2012dq} and Caputo derivative,  we find that the phase transition of Kerr AdS black hole   arises at $\beta=1/3$.

Therefore, near the critical point,  fractional order of the phase transition is no longer $\beta=4/3$. This may suggest that $4/3$ order FPT is not universal and only holds for static black holes with spherical symmetry.

\section{Conclusion}
\label{Conclusion}
To summarize, we have studied the  thermodynamical phase transitions of several AdS black holes prototypes according to the generalized Ehrenfest classification.  By using the Caputo fractional derivatives of thermodynamic potentials for both a charged black hole surrounded by quintessence,  $5D$ Gauss-Bonnet  and RN-AdS$_D$ black holes,  we found  that the fractional derivatives of the Gibbs free energy is always discontinuous at the critical point for $\beta = 4/3$ order, but diverges when $\beta>4/3$. These results suggested that, though the $4/3$  phase transition order always holds as far as one deals with static black holes with spherical symmetry, it is not a universal order since it is altered by the geometric symmetry, as revealed by the analysis of the AdS-Kerr solution.

\bibliographystyle{elsarticle-num}
\bibliography{myreferences} 

\end{document}